\documentclass[aps,prl,reprint,superscriptaddress,nofootinbib,longbibliography]{revtex4-2}

\usepackage{amsmath,amssymb,bm,mathtools,mathrsfs}
\usepackage{graphicx}
\usepackage{microtype}
\usepackage[colorlinks=true,allcolors=blue]{hyperref}
\hypersetup{
 pdftitle={Divergence Geometry of Quantum Multi-Mpemba Effects},
 pdfauthor={Domingos S. P. Salazar}
}
\usepackage{booktabs}

\newcommand{\Tr}{\operatorname{Tr}}
\newcommand{\conv}{\operatorname{conv}}

\newcommand{\dd}{\mathrm{d}}
\newcommand{\cF}{\mathfrak F_2}
\newcommand{\Dl}{\Delta_\lambda}
\newcommand{\Df}{\Delta_f}

\newcommand{\steady}{\omega}

\begin{document}

\title{Divergence Geometry of Quantum Multi-Mpemba Effects}

\author{Domingos S. P. Salazar}
\affiliation{Universidade Federal Rural de Pernambuco, Departamento de F\'isica, Recife, PE, Brazil}

\date{July 30, 2026}

\begin{abstract}
Two quantum states may relax toward the same steady state, yet the one that starts farther away can overtake the closer one. This is the quantum Mpemba effect. The same pair can reverse order more than once, but these crossings may depend on the divergence used to compare them. We ask which crossings persist for every normalized operator-convex Petz divergence. Because each such divergence is a positive average of one-parameter $\chi^2$ kernels, the problem reduces to the sign of one profile across the full kernel range. Alternating sign margins guarantee repeated family-wide reversals, while finite dimension yields a polynomial positivity test. A simple real slow mode fixes the common late-time order; near stationarity, coherence between unequal-eigenvalue sectors makes the order divergence dependent. In a trapped-ion qutrit ideal model, the reported preparation shows divergence-selective crossings. A nearby preparation is a candidate for at least two family-wide reversals.
\end{abstract}

\maketitle

\textit{Introduction.---}
Which of two initial states relaxes faster to a steady state? A single timescale may not settle the question, because the starting state can change the observed order. In the Mpemba effect, a preparation that begins farther from a common stationary state later overtakes a closer one \cite{Mpemba1969,LuRaz2017,Klich2019,Kumar2020}. Quantum studies examine this behavior through decay modes, coherence, symmetry restoration, memory, batteries, random circuits, and engineered resets \cite{Carollo2021,Chatterjee2023,NavaEgger2024,WangWang2024,Moroder2024,Rylands2024,Strachan2025,Medina2025,LiuRandom2024,Qian2025,Turkeshi2025,Bao2025,NavaPontus2025}. Experiments and simulations with trapped ions, other simulators, isolated many-body systems, and superconducting processors explore several of these regimes \cite{Joshi2024,Shapira2024,Zhang2025,AresMixed2025,Yu2025,Yamashika2026,Xu2026}. Recent reviews give broader context \cite{AresReview2025,TezaReview2026}.

Quantum relaxation raises a further question: what counts as closer to stationarity? Several inequivalent divergences can compare the same density operators. We study the normalized operator-convex Petz family, a broad but specified set of divergences \cite{Petz1986,LesniewskiRuskai1999,Hiai2011,HiaiErratum2017}. Up to positive normalization and affine terms that do not affect the ordering, it includes forward and reverse Umegaki relative entropy \cite{Umegaki1962,Petz1986}, their Jeffreys symmetrization \cite{Jeffreys1946,SousaPires2026}, squared Petz--Hellinger divergence \cite{Petz2010,Salazar2025f}, and the continuous family of Petz $\chi^2$ kernels \cite{LesniewskiRuskai1999,Salazar2025f}. For $0<\alpha<1$ and $1<\alpha\le2$, Petz R\'enyi divergences give the same pairwise order as normalized family members after a strictly increasing reparameterization. Trace distance, Hilbert--Schmidt distance, and sandwiched R\'enyi divergences generally fall outside this linear Petz family \cite{Hiai2011,MullerLennert2013}.

Mpemba studies also compare states using quantities outside the Petz-divergence family: entanglement asymmetry, Hilbert--Schmidt and trace distances, ergotropy, and Krylov complexity \cite{YamashikaAres2024,Xia2026,LiErgotropic2025,AlishahihaVasli2026}. For thermal populations, classical thermomajorization supplies a measure-independent criterion \cite{VuHayakawa2025}. Resource-theoretic analyses use free-state orders and Petz--R\'enyi monotones to organize classical and quantum effects \cite{Summer2026}. A driven-granular example shows directly that changing the distance can change the inferred order \cite{Biswas2023Distance}.

Two admissible divergences may reverse the order of the same pair of states, or assign different times and numbers of crossings \cite{Qian2025,Summer2026}. In multi-Mpemba dynamics, fast, intermediate, and slow modes can produce repeated reversals \cite{Chalas2024,Chatterjee2024multi,Xia2026}. A double crossing under one divergence establishes the order only for that divergence. To claim the same behavior across the Petz family, we must check the whole continuum.

We compare two noncommuting relaxation paths with the same reference state. How can we tell whether every divergence gives them the same order? Operator convexity expresses each Petz divergence as a positive mixture of one-parameter kernels \cite{LesniewskiRuskai1999,Hiai2011,HiaiErratum2017}; Ref.~\cite{Salazar2025f} stated this explicitly as a quantum $\chi_\lambda^2$ decomposition. We therefore compare the paths kernel by kernel. If their difference has one sign for every kernel, the whole normalized Petz family agrees on the order. The profile extrema give the full range of divergence differences. Strict alternating signs at selected times force repeated family-wide crossings with positive margins. In finite dimension, the sign check reduces to algebra. We also show when a common simple real slow mode fixes the late-time order, and when coherence between unequal-eigenvalue sectors creates the leading near-equilibrium profile variation. We apply the results to a trapped-ion qutrit ideal model.

\begin{figure*}[t!]
 \includegraphics[width=0.995\textwidth]{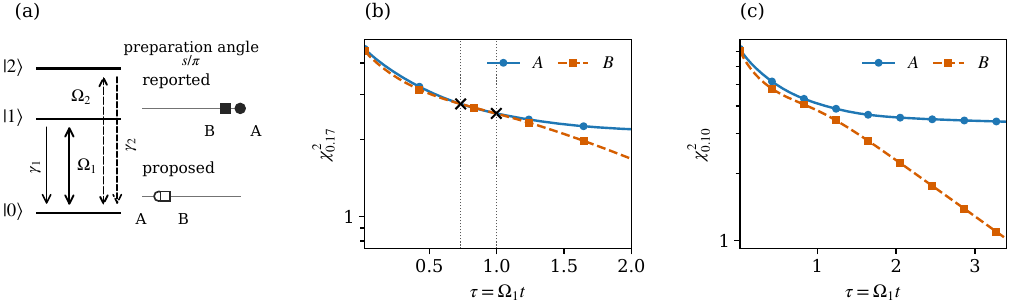}
 \caption{\textbf{Divergence dependence at the reported preparation setting.} Two states, $A$ and $B$, relax toward the same stationary state in the reported trapped-ion ideal model \cite{Xia2026}. In (b) and (c), the horizontal axis is elapsed dimensionless time $\tau:=\Omega_1t$; the vertical axis is the Petz kernel divergence $\chi_\lambda^2$ of each state from stationarity. A lower curve is closer to the stationary state by that divergence. On the plotted windows, the curves cross twice for $\lambda=0.17$ but the numerical scan finds no crossing for $\lambda=0.10$. (a) The qutrit has couplings $\Omega_{1,2}$ and decays $\gamma_{1,2}$; its preparation angle $s$ selects the initial pure state and is plotted as $s/\pi$. Filled markers show the reported pair $(s_A,s_B)=(\pi,0.85\pi)$, while separated open markers show the proposed pair $(0.17\pi,0.235\pi)$. (b) The two crossings for $\lambda=0.17$ exchange which state is closer in the magnified window $0.02\le\tau\le2$. (c) The $\lambda=0.10$ scan covers $0.02\le\tau\le3.4$ and does not exclude roots by interval proof.}
\label{fig:reported}
\end{figure*}

\textit{Universal-order criterion.---}
Let $\rho_A(t)$ and $\rho_B(t)$ be continuous faithful (full-rank) state paths on the time interval of interest. They are compared with the same fixed faithful reference state $\steady$. For faithful states $\rho$ and $\sigma$, define left and right multiplication by $L_\rho(X):=\rho X$ and $R_{\sigma^{-1}}(X):=X\sigma^{-1}$, and set $\Delta_{\rho\vert\sigma}:=L_\rho R_{\sigma^{-1}}$. A real operator-convex function $f$ on $(0,\infty)$ then gives the Petz quasi-entropy $D_f(\rho\Vert\sigma):=\Tr[\sigma^{1/2}f(\Delta_{\rho\vert\sigma})(\sigma^{1/2})]$ \cite{Petz1986,Hiai2011}. Affine terms do not affect the ordering, and an overall positive scale does not change it. We therefore use the class $\cF$ defined by $f(1)=f'(1)=0$ and $f''(1)=2$; the subscript in $\mathfrak F_2$ records this normalization.

\begin{samepage}
\noindent For $\lambda\in[0,1]$, define the kernel generator and its Petz divergence by
\begin{equation}
 f_\lambda(u):=\frac{(u-1)^2}{(1-\lambda)u+\lambda},\qquad
 \chi_\lambda^2(\rho\Vert\sigma):=D_{f_\lambda}(\rho\Vert\sigma).
 \label{eq:kernel}
\end{equation}
\end{samepage}
In Fig.~\ref{fig:reported}, changing $\lambda$ changes the kernel used to compare the same two paths. The reference $\steady$ is the stationary state, and $\tau=\Omega_1t$ is dimensionless time.

Every $f\in\cF$ can be written as a positive mixture of these kernels \cite{Petz1986,LesniewskiRuskai1999,Hiai2011,HiaiErratum2017,Salazar2025f}:
\begin{equation}
 D_f(\rho\Vert\sigma)=\int_{[0,1]}\chi_\lambda^2(\rho\Vert\sigma)\,\dd\mu_f(\lambda),
 \label{eq:mixture}
\end{equation}
Here $\mu_f$ is a probability measure, so Eq.~\eqref{eq:mixture} is an average. Supplemental Material proves the representation, its normalization and endpoint conventions, and the converse from probability measures to normalized generators. We call $\chi_\lambda^2$ the Petz $\chi^2$ kernels of the normalized family. For the two paths, define the kernel ordering profile $\Dl(t):=\chi_\lambda^2(\rho_A(t)\Vert\steady)-\chi_\lambda^2(\rho_B(t)\Vert\steady)$. For a family member, define the corresponding difference $\Df(t):=D_f(\rho_A(t)\Vert\steady)-D_f(\rho_B(t)\Vert\steady)$. A positive difference means that $A$ has a larger divergence from $\steady$ than $B$. Applying Eq.~\eqref{eq:mixture} to both paths shows that the difference for $f$ is the average of the kernel differences:
\begin{equation}
 \Df(t)=\int\Dl(t)\,\dd\mu_f(\lambda).
 \label{eq:delta}
\end{equation}
If every kernel gives the same order, so does every positive average. Conversely, fix $\lambda_0\in[0,1]$ and choose the point-mass measure $\mu_f=\delta_{\lambda_0}$. Equation~\eqref{eq:mixture} then gives $D_f=\chi_{\lambda_0}^2$: this individual kernel is itself a member of the normalized family. These two observations give the exact completeness relation
\begin{equation}
 \Df(t)\le0\ \ \forall f\in\cF
 \quad\Longleftrightarrow\quad
 \Dl(t)\le0\ \ \forall\lambda\in[0,1].
 \label{eq:complete}
\end{equation}
Equation~\eqref{eq:complete} is our first main result: checking the sign of one profile for every $\lambda$ decides the order for the entire normalized Petz family.

The profile also tells us how much the divergences can disagree. Let $m(t):=\min_{\lambda\in[0,1]}\Dl(t)$ and $M(t):=\max_{\lambda\in[0,1]}\Dl(t)$. As $f$ varies over the normalized family, its difference takes every value between these extrema:
\begin{equation}
 \{\Df(t):f\in\cF\}=[m(t),M(t)].
 \label{eq:interval}
\end{equation}
If $m>0$ or $M<0$, every family member gives the same strict order. If $m<0<M$, admissible divergences give opposite orders. The width $M-m$ measures how strongly the order can depend on the divergence.

A crossing compares orders at more than one time. The same divergence must be used at every time, so the same measure $\mu_f$ averages every component. For sampled times $T:=(t_1,\ldots,t_k)$ and $g=f$ or $\lambda$, collect the differences in $\bm\Delta_g(T):=(\Delta_g(t_1),\ldots,\Delta_g(t_k))$. For a set $S\subset\mathbb R^k$, define $\conv S$ as the set of finite convex combinations $\sum_a w_ax_a$ with $x_a\in S$, $w_a\ge0$, and $\sum_aw_a=1$. The mixture now gives
\begin{equation}
 \{\bm\Delta_f(T):f\in\cF\}
 =\conv\{\bm\Delta_\lambda(T):0\le\lambda\le1\}.
 \label{eq:hull}
\end{equation}
As $\lambda$ runs from $0$ to $1$, the vector $\bm\Delta_\lambda(T)$ traces a curve in $\mathbb R^k$. Averaging its points with $\mu_f$ gives the difference vector for one divergence. Equation~\eqref{eq:hull} says that these averages fill the curve's convex hull. For a chosen divergence, the Fenchel--Eggleston theorem shows that at most $k$ kernels can reproduce its vector at these $k$ times \cite{Eggleston1958}. To force crossings for every divergence, we need the required sign at every $\lambda$. Supplemental Material proves the hull identity and shows how uniform profile errors affect crossing margins.

Our second main result is a sufficient condition for repeated family-wide crossings. For two crossings, choose three times $t_0<t_1<t_2$. If every kernel orders path $A$ farther from stationarity, then $B$ farther, then $A$ farther, every divergence must reverse the order twice. More generally, choose $t_0<\cdots<t_k$. A sign $s=\pm1$ allows either starting order. The condition is
\begin{equation}
 s(-1)^j\Dl(t_j)>0\qquad
 \forall\lambda\in[0,1],\quad j=0,\ldots,k,
 \label{eq:sign-anchors}
\end{equation}
The positive average in Eq.~\eqref{eq:delta} preserves each strict anchor sign. Continuity in time then gives at least one zero for every $f\in\cF$ in each interval $(t_{j-1},t_j)$. This sufficient condition yields a family-wide multi-Mpemba effect within $\cF$: the crossing times may differ, but every member crosses in each interval. Define the minimum anchor margin as $\eta:=\min_{j,\lambda}s(-1)^j\Delta_\lambda(t_j)>0$. The conclusion survives any uniform profile error smaller than $\eta$. A common crossing instant for all divergences is not required and would be nongeneric.

\textit{Finite sign tests.---}
Equation~\eqref{eq:complete} asks for a sign check at every $\lambda$. In finite dimension, that continuum test can be reduced to algebra. Write $\rho=\sum_ip_i|p_i\rangle\langle p_i|$ and $\steady=\sum_jq_j|q_j\rangle\langle q_j|$. The Nussbaum--Szko\l{}a reduction \cite{NussbaumSzkola2009,AndroulakisJohn2024} gives
\begin{equation}
 \chi_\lambda^2(\rho\Vert\steady)=
 \sum_{ij}\frac{(p_i-q_j)^2|\langle p_i|q_j\rangle|^2}
 {(1-\lambda)p_i+\lambda q_j}.
 \label{eq:NS}
\end{equation}
Set $z:=\lambda/(1-\lambda)$, which maps $0\le\lambda<1$ to $z\ge0$. For each term in Eq.~\eqref{eq:NS}, put $r_{ij}:=p_i/q_j$ and $c_{ij}:=q_j(r_{ij}-1)^2|\langle p_i|q_j\rangle|^2\ge0$. Dividing by $1+z$ gives
\begin{align*}
 F_{\rho|\steady}(z)
 &:=\frac{\chi^2_{z/(1+z)}(\rho\Vert\steady)}{1+z}\\
 &=\sum_{ij}\frac{c_{ij}}{r_{ij}+z}.
\end{align*}
This is a Stieltjes transform of a positive measure \cite{Akhiezer1965}. In dimension $d$, it has at most $d^2$ poles. At a fixed time, write $F_{X,t}(z):=F_{\rho_X(t)|\steady}(z)$ for path $X=A,B$, and use matching superscripts on its spectral data. Subtracting the transforms gives the promised sign test:
\begin{align*}
 \frac{\Delta_{z/(1+z)}(t)}{1+z}
 &=F_{A,t}(z)-F_{B,t}(z)\\
 &=\sum_{ij}\frac{c^A_{ij}(t)}{r^A_{ij}(t)+z}\\
 &\quad-\sum_{ij}\frac{c^B_{ij}(t)}{r^B_{ij}(t)+z}\\
 &=\frac{P_t(z)}{Q_t(z)}.
\end{align*}
All $r^X_{ij}(t)$ are positive, so the product denominator $Q_t(z)$ can be chosen positive for $z\ge0$. Since $1+z>0$, the sign of the original profile is the sign of $P_t(z)$. Its degree is at most $2d^2-1$, or $17$ for a qutrit. Family-wide ordering thus becomes a polynomial sign test on $z\ge0$, with a separate check at $\lambda=1$. Applying this test at each anchor time can establish the signs required by Eq.~\eqref{eq:sign-anchors}. Exact rational or algebraic input permits Sturm or subresultant root isolation \cite{BasuPollackRoy2006}. Supplemental Material gives the rational reduction, the endpoint check, and derivative-controlled grid bounds.

\textit{Late-time order and divergence dependence.---}
At late times, can the choice of divergence still change the order? We answer this under a specific dynamical assumption. Both paths evolve under the same finite-dimensional CPTP semigroup $\mathcal E_t:=e^{t\mathcal L}$ with faithful stationary state $\steady$. The leading nonstationary eigenvalue contributing to either path is assumed to be a simple real value $-\kappa_1<0$. Its Hermitian right eigenoperator $R_1\ne0$ satisfies $\mathcal L(R_1)=-\kappa_1R_1$, and its real part is separated from the rest of the contributing spectrum. For $X\in\{A,B\}$, let $a_X\in\mathbb R$ be the corresponding amplitude and assume $a_A^2\ne a_B^2$. Then
\begin{equation}
 \rho_X(t)=\steady+a_Xe^{-\kappa_1t}R_1+o(e^{-\kappa_1t}).
 \label{eq:slow}
\end{equation}
The little-$o$ term in Eq.~\eqref{eq:slow} is measured in any fixed finite-dimensional matrix norm. Let $G_f^\steady$ be the positive Hessian quadratic form of $D_f$ at $\steady$, and set $G_\lambda^\steady:=G_{f_\lambda}^\steady$ for the extremal kernels. Uniformly over $f\in\cF$, we obtain
\begin{equation}
 \Df(t)=\bigl(a_A^2-a_B^2\bigr)e^{-2\kappa_1t}
 G_f^\steady(R_1)+o(e^{-2\kappa_1t}).
 \label{eq:tail}
\end{equation}
The scalar remainder is also uniform over $f\in\cF$. Because $\min_{\lambda\in[0,1]}G_\lambda^\steady(R_1)>0$, the sign at sufficiently late times is the same for every normalized Petz divergence under these assumptions. If $a_A^2=a_B^2$, higher-order modes and higher-order divergence terms decide the asymptotic order instead. Supplemental Material proves the uniform local remainder and the slow-mode expansion.

What makes two Petz divergences disagree near stationarity? Work in the eigenbasis $\steady=\sum_i\steady_i|i\rangle\langle i|$ and write $X_{ij}:=\langle i|X|j\rangle$. For $\rho=\steady+\epsilon X$ with $X=X^\dagger$, $\Tr X=0$, and $|\epsilon|$ small enough that $\rho$ is faithful,
\begin{align}
 \chi_\lambda^2(\steady+\epsilon X\Vert\steady)
 &=\epsilon^2G_\lambda^\steady(X)+O(\epsilon^3),
 \label{eq:localexpansion}\\
 G_\lambda^\steady(X)
 &=\sum_{ij}\frac{|X_{ij}|^2}{(1-\lambda)\steady_i+\lambda\steady_j}.
 \label{eq:hessian}
\end{align}
The diagonal contribution is $\sum_i|X_{ii}|^2/\steady_i$, independent of $\lambda$. At quadratic order, only coherences between stationary-state sectors with unequal eigenvalues make the profile depend on $\lambda$. Supplemental Material derives the uniform expansion. It also suggests a near-stationary experimental control: use tomography to compare each state with a version dephased in the eigenbasis of $\steady$, then compare the profile spread $M-m$ before and after removing the off-diagonal terms. If that spread vanishes after dephasing, coherence accounts for the divergence dependence at quadratic order, subject to the $O(\epsilon^3)$ remainder and reconstruction error. This test concerns divergence dependence. Whether one path overtakes the other still depends on trajectory and mode competition; away from stationarity, population terms can restore profile variation.

\begin{figure*}[t!]
 \includegraphics[width=0.995\textwidth]{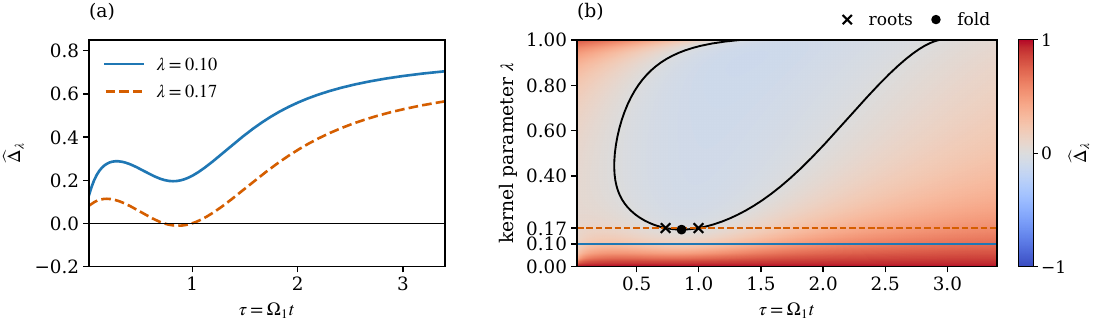}
 \caption{\textbf{Reported-setting divergence cuts and kernel topology.} The ideal-model calculation uses $(s_A,s_B)=(\pi,0.85\pi)$ \cite{Xia2026}; $\tau=\Omega_1t$ begins at the pure preparation. The compressed difference is $\widehat\Delta_\lambda:=\Delta_\lambda/(1+|\Delta_\lambda|)$. (a) The solid cut $\widehat\Delta_{0.10}$ stays above zero in the plotted window, while the dashed cut $\widehat\Delta_{0.17}$ crosses zero twice. Thus the crossings depend on the chosen kernel divergence. (b) Shading shows the directly evaluated grid of $\widehat\Delta_\lambda$ for $0\le\lambda\le1$ and $0.02\le\tau\le3.4$. The zero frontier is the unbroken curve. Horizontal guides locate the two cuts; crosses mark the independently solved $\lambda=0.17$ crossings. The dot marks the numerical merger candidate $(\lambda_c,\tau_c)=(1.6320\times10^{-1},\,8.6095\times10^{-1})$. Signed colorbar ticks, line styles, and the zero frontier make the encoding readable without color alone. Both panels use ideal-model trajectories.}
 \label{fig:selective}
\end{figure*}

\begin{figure*}[t!]
 \includegraphics[width=0.995\textwidth]{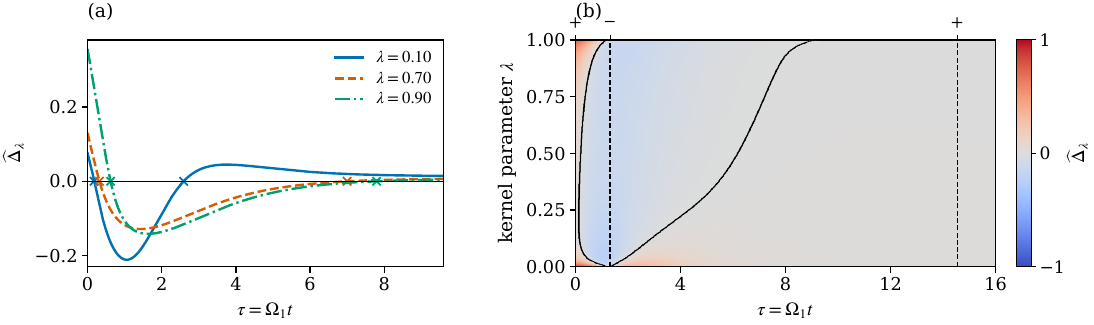}
 \caption{\textbf{Proposed-setting candidate for at least two family-wide order reversals.} Both preparations, $(s_A,s_B)=(0.17\pi,0.235\pi)$, evolve under the reported ideal generator for $t_{\rm pre}=0.04/\Omega_1$. The resulting full-rank floating-point states set $\tau=0$. (a) Direct compressed cuts are shown for $\lambda=0.10$ (solid), $0.70$ (dashed), and $0.90$ (dash-dotted). Crosses mark independently solved roots. These three cuts illustrate the behavior; they do not establish the family-wide claim. (b) The directly evaluated grid covers $0\le\lambda\le1$ and $0\le\tau\le16$. Black markers show independently solved sign-changing roots. Floating-point estimates of the optimized profile extrema at $\tau=0$, $1.32$, and $14.54$ give the sign sequence $+,-,+$. If these signs hold over the entire kernel interval, Eq.~\eqref{eq:sign-anchors} gives at least two crossings for every normalized Petz divergence. Thus this new preparation is a numerical candidate for crossings throughout the whole family.}
 \label{fig:universal}
\end{figure*}

\textit{Application: Trapped-ion test.---}
We test the criterion in the ideal model of the driven-dissipative $^{40}\mathrm{Ca}^{+}$ qutrit used in the recent multi-Mpemba experiment \cite{Xia2026}. With $\Omega_1=1$, the reported generator is
\begin{align}
 H&=\sum_{\alpha=1}^2\frac{\Omega_\alpha}{2}
 (|0\rangle\langle\alpha|+|\alpha\rangle\langle0|),
 \label{eq:model-H}\\
 J_\alpha&=\sqrt{\gamma_\alpha}|0\rangle\langle\alpha|,
 \nonumber\\[-2pt]
 \Omega_2&=6.0\times10^{-2},\qquad
 \gamma_1=2.0,\qquad
 \gamma_2=1.5\times10^{-3}.
 \label{eq:model}
\end{align}
These operators define the GKSL Liouvillian $\mathcal L$ \cite{Gorini1976,Lindblad1976}; its explicit form is in Supplemental Material. The initial states are $|s\rangle:=\cos s|\phi_-\rangle-i\sin s|\phi_+\rangle$. Here $|\phi_\pm\rangle$ are selected from the Hermitian left slow mode, with their components and phase convention given in Supplemental Material.

For the reported setting, we use this generator and $(s_A,s_B)=(\pi,0.85\pi)$. Figures~\ref{fig:reported} and \ref{fig:selective} show synthetic ideal-generator trajectories, not reprocessed experimental state data. At positive evolved times, both states are faithful. In the displayed window, the numerical kernel field has zero branches ending near a merger candidate: only part of the normalized family shows two crossings. This supports a Petz-divergence-selective reading of the ideal model. A continuum proof and propagation of experimental state uncertainty remain to be done.

A search over the same preparation family gives the floating-point candidate in Fig.~\ref{fig:universal}. Let $\mathcal E_t=e^{t\mathcal L}$ be evolution under the reported generator, and start from $\rho_X^{(0)}:=|s_X\rangle\langle s_X|$ with $(s_A,s_B)=(0.17\pi,0.235\pi)$. Evolve both states for the same elapsed time $t_{\rm pre}=0.04/\Omega_1$. This gives $\bar\rho_X:=\mathcal E_{t_{\rm pre}}(\rho_X^{(0)})$. The minimum floating-point eigenvalues of $\bar\rho_A$ and $\bar\rho_B$ are $1.888\times10^{-7}$ and $1.678\times10^{-7}$, respectively. Both exceed the recorded $10^{-12}$ positivity tolerance. The comparison then starts at $\tau=0$ with $\rho_X(\tau):=\mathcal E_{\tau/\Omega_1}(\bar\rho_X)$, after the common pre-evolution.

Figure~\ref{fig:universal} shows three representative kernel cuts and a numerical field over $0\le\lambda\le1$. The family-wide candidate rests on optimized floating-point extrema, not on the three cuts alone.

The floating-point search gives $\widetilde m(0)=5.069\times10^{-2}$, $\widetilde M(1.32)=-5.237\times10^{-2}$, and $\widetilde m(14.54)=4.705\times10^{-3}$ at presentation precision. These estimated signs, $+,-,+$, support a candidate for at least two order reversals for every normalized Petz divergence. Supplemental Material gives the higher-precision recorded values. The common negative window is narrow, $1.22\lesssim\tau\lesssim1.39$. The last common sign change appears near $\tau\simeq9.18$ in this calculation and joins the slow-mode tail.

The candidate may have more than two crossings for some kernels. At $\lambda=0$, four floating-point roots occur near $\tau=1.33\times10^{-2}$, $6.29\times10^{-1}$, $7.04\times10^{-1}$, and $1.99$; full solver precision is in Supplemental Material. The additional pair survives only in a narrow near-endpoint layer. In this floating-point candidate, the two family-wide reversals are therefore a lower bound for some kernels, not a claim that every kernel has exactly two crossings. Supplemental Material records the rank convention, spectra, solver outputs, retained arrays, and optimization procedure. A family-wide conclusion still requires outward-rounded global bounds.

\textit{Discussion and outlook.---}
The settings make different claims. In the numerical field of Fig.~\ref{fig:selective}, zero branches occupy part of the kernel interval and end near a merger candidate: crossings depend on the divergence. In Fig.~\ref{fig:universal}, estimated extrema at three anchor times have full-interval signs $+,-,+$; a narrow layer near $\lambda=0$ has an additional pair of roots. Under faithfulness and profile continuity, the strict inequalities in Eq.~\eqref{eq:sign-anchors} keep positive margins stable against sufficiently small perturbations.

This criterion complements thermomajorization and resource-theoretic orders \cite{VuHayakawa2025,Summer2026,Jin2026}. It covers the normalized Petz family; trace distance, Hilbert--Schmidt distance, sandwiched R\'enyi divergences, and arbitrary operational tasks generally lie outside it \cite{Hiai2011,MullerLennert2013}. Equation~\eqref{eq:complete} checks the full kernel interval; a finite list of divergences covers only its chosen members. Thermomajorization ranges over all classical monotone measures, where slow-mode amplitude ordering need not suffice \cite{VuHayakawa2025}. Within the Petz family, the common quadratic local structure makes unequal squared amplitudes sufficient for late-time agreement under the simple-mode hypotheses of Eq.~\eqref{eq:tail}.

Tomography could reconstruct $\rho_A(t)$, $\rho_B(t)$, and $\steady$ and evaluate the profile with Eq.~\eqref{eq:NS}, following Ref.~\cite{Xia2026}. A simultaneous state-confidence region could yield a uniform profile band, whose sign could be tested by polynomial root isolation or the grid test in Supplemental Material. The small spectral floor of the proposed pre-evolved states makes error propagation sensitive, so the profile margin alone cannot set measurement precision. Supplemental Material gives a norm-explicit state-to-profile bound.

The ordering and crossing statements in Eqs.~\eqref{eq:mixture}--\eqref{eq:sign-anchors} need continuous state paths and a faithful reference, but allow driven or non-Markovian paths. The late-time theorem additionally needs Markovian spectral structure. A rank-deficient preparation needs a positivity-improving time or sign-stable regularization. One robust-control goal is to maximize the smallest sign-anchor margin; Supplemental Material gives this max--min objective.

One sign profile is enough to decide family-wide order within the normalized Petz class. Its extrema show the possible spread across divergences, and finite-dimensional algebra offers a way to validate its sign. In the numerical trapped-ion ideal model, the reported preparation shows selective crossings. A nearby preparation is a candidate for at least two family-wide reversals; outward-rounded continuum bounds are still needed.

\begin{acknowledgments}
OpenAI Codex (GPT-5) and ChatGPT (Oracle mode; model version not exposed) assisted with literature organization, derivation checking, code debugging, numerical cross-checks, figure-layout iteration, and prose revision. The author specified the scientific constraints and independently reviewed the derivations, citations, simulations, numerical outputs, figures, and generated text, and takes full responsibility for the manuscript.
\end{acknowledgments}

\textit{Data availability.---}The numerical data and code supporting the figures are available from the author upon reasonable request.

\bibliography{references}

\clearpage
\onecolumngrid
\section*{Supplemental Material}

\section{Normalized Petz family and kernel completeness}
\label{SM:kernel}

For faithful density operators $\rho$ and $\sigma$, let $L_\rho(X):=\rho X$,
$R_{\sigma^{-1}}(X):=X\sigma^{-1}$, and
$\Delta_{\rho|\sigma}:=L_\rho R_{\sigma^{-1}}$ act on the Hilbert--Schmidt
operator space. The Petz quasi-entropy generated by a real
operator-convex function $f$ on $(0,\infty)$ is
\begin{equation}
 D_f(\rho\Vert\sigma):=
 \Tr\!\left[\sigma^{1/2}f(\Delta_{\rho|\sigma})(\sigma^{1/2})\right].
 \label{SM(petzdef)}
\end{equation}
For an affine function $\ell(u):=a+b(u-1)$, cyclicity of the trace gives
\begin{equation}
 D_\ell(\rho\Vert\sigma)
 =a\,\Tr\sigma+b\bigl(\Tr\rho-\Tr\sigma\bigr).
 \label{SM(affine)}
\end{equation}
Thus affine terms cancel from the difference between two equal-trace
states evaluated against the same reference. We may impose
$f(1)=f'(1)=0$ without changing any ordering statement in the Letter.

We first turn the operator-convex representation into a probability
average of kernels. The standard representation on the positive
half-line gives $b_0,b_1\ge0$ and a positive measure $\nu_f$ on
$(0,\infty)$ with $\int(1+s)^{-1}\dd\nu_f(s)<\infty$ such that
\begin{equation}
 \begin{split}
 f(u)={}&f(1)+f'(1)(u-1)+b_1(u-1)^2
 +b_0\frac{(u-1)^2}{u}\\
 &+\int_{(0,\infty)}\frac{(u-1)^2}{u+s}\,\dd\nu_f(s).
 \end{split}
 \label{SM(rawrepresentation)}
\end{equation}
This representation is a known result
\cite{LesniewskiRuskai1999,Hiai2011,HiaiErratum2017}.
Set $\lambda(s):=s/(1+s)$ and define a measure on the closed kernel
interval by
\begin{equation}
 \mu_f:=b_0\delta_0+b_1\delta_1+
 \lambda_\#\!\left(\frac{\nu_f(\dd s)}{1+s}\right),
 \label{SM(pushforward)}
\end{equation}
where $\lambda_\#$ denotes pushforward. Since
$(1-\lambda(s))u+\lambda(s)=(u+s)/(1+s)$,
Eq.~\eqref{SM(rawrepresentation)} becomes
\begin{align}
 f(u)-f(1)-f'(1)(u-1)
 &=\int_{[0,1]}f_\lambda(u)\,\dd\mu_f(\lambda),
 \label{SM(fmixture)}\\
 f_\lambda(u)&=\frac{(u-1)^2}{(1-\lambda)u+\lambda},
 \qquad 0\le\lambda\le1.
 \label{SM(kernelgenerator)}
\end{align}
The endpoint atoms are the terms $(u-1)^2/u$ at $\lambda=0$ and
$(u-1)^2$ at $\lambda=1$. Writing $u=1+x$ gives
$f_\lambda(1+x)=x^2/[1+(1-\lambda)x]$; hence
$f_\lambda''(1)=2$ uniformly in $\lambda$. Differentiating the finite
measure representation at $u=1$ yields
\begin{equation}
 \mu_f([0,1])=\frac{f''(1)}{2}.
 \label{SM(mass)}
\end{equation}
Thus $f''(1)=2$ exactly when $\mu_f$ has unit mass. In that case
$\mu_f$ is a probability measure. The converse matters for the order
criterion: every probability measure on $[0,1]$ defines a normalized
operator-convex function through Eq.~\eqref{SM(fmixture)}. Positive
integration preserves operator convexity, and the derivatives at $u=1$
are uniform in $\lambda$.

The relation to Petz R\'enyi divergences is explicit. For
$0<\alpha<1$ and $1<\alpha\le2$, respectively, define
\begin{align}
 f_{\alpha,<}(u)
 &=\frac{2}{\alpha(1-\alpha)}
   \bigl[1-u^\alpha+\alpha(u-1)\bigr],
 \label{SM(renyigeneratorless)}\\
 f_{\alpha,>}(u)
 &=\frac{2}{\alpha(\alpha-1)}
   \bigl[u^\alpha-1-\alpha(u-1)\bigr].
 \label{SM(renyigeneratorgreater)}
\end{align}
Both generators satisfy $f(1)=f'(1)=0$ and $f''(1)=2$. For density
operators, the Petz R\'enyi divergences \cite{Hiai2011} obey
\begin{align}
 D_\alpha^{\rm P}(\rho\Vert\sigma)
 &=-\frac{1}{1-\alpha}
 \log\!\left[
 1-\frac{\alpha(1-\alpha)}{2}
 D_{f_{\alpha,<}}(\rho\Vert\sigma)
 \right],
 &&0<\alpha<1,
 \label{SM(renyireparamless)}\\
 D_\alpha^{\rm P}(\rho\Vert\sigma)
 &=\frac{1}{\alpha-1}
 \log\!\left[
 1+\frac{\alpha(\alpha-1)}{2}
 D_{f_{\alpha,>}}(\rho\Vert\sigma)
 \right],
 &&1<\alpha\le2.
 \label{SM(renyireparamgreater)}
\end{align}
The right-hand sides are strictly increasing functions of the corresponding
normalized Petz divergences, which proves the pairwise-ordering statement
used in the main text.

We now pass from the generator identity to the divergence identity.
Diagonalize
$\rho=\sum_i p_i|p_i\rangle\langle p_i|$ and
$\sigma=\sum_jq_j|q_j\rangle\langle q_j|$. The matrix units
$E_{ij}=|p_i\rangle\langle q_j|$ satisfy
\begin{equation}
 \Delta_{\rho|\sigma}(E_{ij})=\frac{p_i}{q_j}E_{ij},
 \qquad
 \left|\langle E_{ij},\sigma^{1/2}\rangle_{\rm HS}\right|^2
 =q_j|\langle p_i|q_j\rangle|^2.
 \label{SM(modulareigen)}
\end{equation}
Because the spectrum is finite, Eq.~\eqref{SM(fmixture)} applies
eigenvalue by eigenvalue. We may interchange the finite sum, integral,
and trace. Equations~\eqref{SM(petzdef)} and \eqref{SM(modulareigen)}
then give
\begin{equation}
 D_f(\rho\Vert\sigma)=
 \sum_{ij}q_j f(p_i/q_j)|\langle p_i|q_j\rangle|^2
 =\int_{[0,1]}\chi_\lambda^2(\rho\Vert\sigma)\,
 \dd\mu_f(\lambda),
 \label{SM(functionalcalculus)}
\end{equation}
which proves Eq.~\eqref{eq:mixture}. For $f=f_\lambda$ this also gives
Eq.~\eqref{eq:NS}; the equality with the associated classical
Nussbaum--Szko\l{}a $f$-divergence is standard
\cite{NussbaumSzkola2009,AndroulakisJohn2024}.

Faithfulness gives $p_i,q_j>0$. Every denominator in
Eq.~\eqref{eq:NS} is therefore positive for $0\le\lambda\le1$,
including at both endpoints. The finite sum is continuous on that
closed interval. This argument does not depend on a choice of basis or
on continuous eigenvectors at degeneracies. On a compact time interval,
continuity of the state paths and faithful reference gives continuity
in time through finite-dimensional functional calculus.

At a fixed time, write $g(\lambda):=\Delta_\lambda(t)$. If $g\le0$
at every $\lambda$, its average under any probability measure is
nonpositive. Conversely, if $g(\lambda_0)>0$, choose the point mass
$\delta_{\lambda_0}$. It gives the admissible normalized generator
$f_{\lambda_0}$ and a positive divergence difference. Hence
\begin{equation}
 \int g(\lambda)\,\dd\mu(\lambda)\le0\quad
 \forall\mu\in\mathcal P([0,1])
 \quad\Longleftrightarrow\quad
 g(\lambda)\le0\quad\forall\lambda.
 \label{SM(completenessproof)}
\end{equation}
Continuity on the compact kernel interval ensures that $g$ reaches
its minimum and maximum. Point masses at those locations give the two
extreme averages. Mixtures of the point masses give every value
between them. Thus
\begin{equation}
 \left\{\int g\,\dd\mu:\mu\in\mathcal P([0,1])\right\}
 =\conv g([0,1])=[\min g,\max g].
 \label{SM(intervalproof)}
\end{equation}

\section{Multi-time convex geometry and robust crossings}
\label{SM:multitime}

For $T=(t_1,\ldots,t_k)$, integration is componentwise:
\begin{equation}
 (\Delta_f(t_1),\ldots,\Delta_f(t_k))
 =\int_{[0,1]}\bm\Delta_\lambda(T)\,\dd\mu_f(\lambda).
 \label{SM(vectorintegral)}
\end{equation}
The map $\lambda\mapsto\bm\Delta_\lambda(T)$ is continuous, so its image is compact and connected. The integral in Eq.~\eqref{SM(vectorintegral)} lies in the convex hull of that image. Conversely, an atomic probability measure realizes every finite convex combination. This proves Eq.~\eqref{eq:hull}. The Fenchel--Eggleston theorem \cite{Eggleston1958} further shows that any point in the hull has the form
\begin{equation}
 \sum_{a=1}^{k}w_a\bm\Delta_{\lambda_a}(T),\qquad
 w_a\ge0,\quad\sum_aw_a=1.
 \label{SM(fenchel-eggleston)}
\end{equation}
Thus at most $k$ Petz kernels reproduce any admissible difference vector at $k$ sampled times.

Suppose Eq.~\eqref{eq:sign-anchors} holds with margins $\varepsilon_j>0$. For every $f\in\cF$,
\begin{equation}
 s(-1)^j\Delta_f(t_j)=\int s(-1)^j\Delta_\lambda(t_j)\,\dd\mu_f(\lambda)\ge\varepsilon_j.
 \label{SM(margintransfer)}
\end{equation}
The sign changes between successive anchor times. Continuity in time therefore forces at least one zero in each $(t_{j-1},t_j)$. Suppose a reconstructed profile $\widetilde\Delta_\lambda(t_j)$ obeys
\begin{equation}
 \sup_{\lambda\in[0,1]}|\widetilde\Delta_\lambda(t_j)-\Delta_\lambda(t_j)|<\varepsilon_j,
 \label{SM(errorrobust)}
\end{equation}
Then the reconstructed profile keeps the same sign at every anchor, and hence the lower bound of one crossing in each anchor interval is unchanged.

For preparation controls $\vartheta$ in an admissible set $\Theta$, let
$m_\vartheta(t)$ and $M_\vartheta(t)$ be the corresponding profile extrema.
Over an observation window $[0,T]$, define the robust double-reversal margin
\begin{equation}
 \begin{aligned}
 \mathcal R_2(\Theta,T)
 &:=\sup_{\vartheta\in\Theta}\ \sup_{0\le t_0<t_1<t_2\le T}\\[-2pt]
 &\quad\max\Bigl\{
 \min\!\bigl[m_\vartheta(t_0),-M_\vartheta(t_1),
                  m_\vartheta(t_2)\bigr],\\[-2pt]
 &\hspace{58pt}
 \min\!\bigl[-M_\vartheta(t_0),m_\vartheta(t_1),
                  -M_\vartheta(t_2)\bigr]
 \Bigr\}.
 \end{aligned}
 \label{SM(robustness)}
\end{equation}
A positive value supplies one of the two alternating sign orientations
with a common positive margin. This is the error tolerance of that
sign-anchor test. The suprema need not be attained unless the control
domain has suitable compactness and the control and time dependence
is continuous.

\section{Stieltjes representation and finite sign tests}
\label{SM:finite-sign}

Let $\rho=\sum_ip_i|p_i\rangle\langle p_i|$ and $\sigma=\sum_jq_j|q_j\rangle\langle q_j|$. Define the Nussbaum--Szko\l{}a likelihood-ratio measure
\begin{equation}
 \nu_{\rho|\sigma}:=\sum_{ij}q_j|\langle p_i|q_j\rangle|^2\,\delta_{p_i/q_j},
 \label{SM(NSmeasure)}
\end{equation}
and $\dd\omega_{\rho|\sigma}(r):=(r-1)^2\dd\nu_{\rho|\sigma}(r)$. Equation~\eqref{eq:NS} becomes
\begin{equation}
 \chi_\lambda^2(\rho\Vert\sigma)=
 \int_0^\infty\frac{(r-1)^2}{(1-\lambda)r+\lambda}\,\dd\nu_{\rho|\sigma}(r).
 \label{SM(NSintegral)}
\end{equation}
With $z:=\lambda/(1-\lambda)$,
\begin{equation}
 \mathscr S_{\rho|\sigma}(z):=\frac{1}{1+z}\chi_{z/(1+z)}^2(\rho\Vert\sigma)
 =\int_0^\infty\frac{\dd\omega_{\rho|\sigma}(r)}{r+z}.
 \label{SM(stieltjesproof)}
\end{equation}
This is a Stieltjes transform of a positive measure \cite{Akhiezer1965}. In dimension $d$, that measure has at most $d^2$ distinct support points. For two states $A,B$, let $n_A$ and $n_B$ count the distinct support points of $\omega_{A|\steady}$ and $\omega_{B|\steady}$. Then
\begin{equation}
 \mathscr S_{A|\steady}(z)-\mathscr S_{B|\steady}(z)=\frac{P_t(z)}{Q_t(z)},
 \qquad Q_t(z)>0\quad(z\ge0),
 \label{SM(rationaldifference)}
\end{equation}
Here $\deg P_t\le n_A+n_B-1\le2d^2-1$. For $0\le\lambda<1$, the kernel order is the sign of $P_t$ on $[0,\infty)$. The endpoint $\lambda=1$ needs its own check, either directly from Eq.~\eqref{eq:NS} or from the leading asymptotic coefficient as $z\to\infty$. Sturm sequences or interval root isolation, together with that endpoint check, give finite machine-checkable sign tests \cite{BasuPollackRoy2006}.

A grid can provide a simpler sign test when a derivative bound is available. Differentiating Eq.~\eqref{eq:NS} gives
\begin{equation}
 \partial_\lambda\chi_\lambda^2(\rho\Vert\steady)=
 \sum_{ij}\frac{(p_i-q_j)^3|\langle p_i|q_j\rangle|^2}
 {[(1-\lambda)p_i+\lambda q_j]^2}.
 \label{SM(profilederivative)}
\end{equation}
Let $L_t\ge\sup_\lambda|\partial_\lambda\Delta_\lambda(t)|$, and evaluate the profile on a grid of mesh $h$. Every $\lambda$ lies within $h/2$ of a grid point, so
\begin{align}
 \min_\lambda\Delta_\lambda(t)&\ge\min_{\lambda_a}\Delta_{\lambda_a}(t)-\frac{L_th}{2},
 \label{SM(gridlower)}\\
 \max_\lambda\Delta_\lambda(t)&\le\max_{\lambda_a}\Delta_{\lambda_a}(t)+\frac{L_th}{2}.
 \label{SM(gridupper)}
\end{align}
If reconstruction adds a separately established uniform error bound $\eta_t$, both bounds also need an $\eta_t$ penalty. This gives a finite-data version of Eq.~\eqref{eq:sign-anchors} without finding every zero of $P_t$.

To carry uncertainty from reconstructed states to the profile, let
$L_\rho(X):=\rho X$ and $R_\steady(X):=X\steady$, and define
\begin{equation}
 K_\lambda(\rho,\steady):=(1-\lambda)L_\rho+\lambda R_\steady,
 \qquad \delta:=\rho-\steady,
 \label{SM(resolventoperator)}
\end{equation}
so that
$\chi_\lambda^2(\rho\Vert\steady)
=\langle\delta,K_\lambda(\rho,\steady)^{-1}\delta\rangle_{\rm HS}$.
Suppose that $\rho,\steady,\rho',\steady'\succeq\eta I$ and let
$B:=\max\{\|\rho-\steady\|_2,\|\rho'-\steady'\|_2\}$, where
$\|\cdot\|_2$ is the Hilbert--Schmidt norm. The resolvent identity gives
\begin{align}
 &\left|
 \chi_\lambda^2(\rho'\Vert\steady')
 -\chi_\lambda^2(\rho\Vert\steady)
 \right|\nonumber\\
 &\quad\le
 \frac{2B}{\eta}
 \|(\rho'-\steady')-(\rho-\steady)\|_2
 {}+
 \frac{B^2}{\eta^2}
 \|K_\lambda(\rho',\steady')-K_\lambda(\rho,\steady)\|_{2\to2}.
 \label{SM(profileconditioning)}
\end{align}
Moreover,
\begin{equation}
 \begin{split}
 &\|K_\lambda(\rho',\steady')-K_\lambda(\rho,\steady)\|_{2\to2}\\
 &\quad\le
 (1-\lambda)\|\rho'-\rho\|_\infty
 +\lambda\|\steady'-\steady\|_\infty .
 \end{split}
 \label{SM(Kperturbation)}
\end{equation}
Indeed,
$K_\lambda(\rho',\steady')-K_\lambda(\rho,\steady)
=(1-\lambda)L_{\rho'-\rho}+\lambda R_{\steady'-\steady}$,
with $\|L_A\|_{2\to2}=\|R_A\|_{2\to2}=\|A\|_\infty$.
The bound holds uniformly for $\lambda\in[0,1]$. Apply it to each
trajectory using a simultaneous region for the estimated states. The
result is a uniform error band for $\Delta_\lambda$. Because the bound
contains inverse powers of $\eta$, the spectral floor affects the
precision needed in an experiment.

\section{Slow-mode universality and coherence selectivity}
\label{SM:dynamics}

Let $\rho=\steady+\epsilon X$ with $X=X^\dagger$, $\Tr X=0$, and $|\epsilon|$ small enough that $\rho$ is faithful. Expanding each extremal kernel uniformly in $\lambda$ gives
\begin{equation}
 \chi_\lambda^2(\steady+\epsilon X\Vert\steady)
 =\epsilon^2G_\lambda^\steady(X)+O(\epsilon^3),
 \label{SM(localexpansion)}
\end{equation}
where, in the eigenbasis of $\steady$,
\begin{equation}
 G_\lambda^\steady(X)=\sum_{ij}\frac{|X_{ij}|^2}{(1-\lambda)\steady_i+\lambda\steady_j}.
 \label{SM(kernelmetric)}
\end{equation}
For the diagonal part $X_{\rm diag}:=\sum_iX_{ii}|i\rangle\langle i|$,
\begin{equation}
 G_\lambda^\steady(X_{\rm diag})=\sum_i\frac{|X_{ii}|^2}{\steady_i},
 \label{SM(populationmetric)}
\end{equation}
which has no $\lambda$ dependence. For a Hermitian coherence pair $i<j$, the coefficient of $|X_{ij}|^2$ is
\begin{equation}
 g_{ij}(\lambda):=
 \frac{\steady_i+\steady_j}{\steady_i\steady_j+\lambda(1-\lambda)(\steady_i-\steady_j)^2}.
 \label{SM(coherencecoefficient)}
\end{equation}
Thus quadratic profile dependence arises only from coherences connecting eigenspaces of $\steady$ with unequal eigenvalues.

Assume Eq.~\eqref{eq:slow} under the semigroup and spectral hypotheses stated in the main text. Applying Eq.~\eqref{SM(localexpansion)} uniformly on the compact kernel interval yields
\begin{equation}
 \Delta_\lambda(t)=\bigl(|a_A|^2-|a_B|^2\bigr)e^{-2\kappa_1t}G_\lambda^\steady(R_1)
 +o(e^{-2\kappa_1t}),
 \label{SM(kerneltail)}
\end{equation}
Here $\min_{\lambda\in[0,1]}G_\lambda^\steady(R_1)>0$. When the squared amplitudes differ, the leading term has one sign for all $\lambda$ at sufficiently late times. Integrating over $\lambda$ proves Eq.~\eqref{eq:tail}. Degenerate slow spaces, complex leading pairs, Jordan blocks, or equal slow amplitudes require separate treatment; asymptotic divergence dependence may survive in those cases.

\section{Qutrit ideal-model calculation and numerical checks}
\label{SM:qutrit}

The trapped-ion study uses the Hilbert--Schmidt norm as its principal
distance and also compares trace distance \cite{Xia2026}. Here we use
its reported ideal generator and preparation parameters. We do not
reprocess experimental density matrices or uncertainty samples.

The two pairs use different time origins. For the reported pair,
$\tau=\Omega_1t$ starts at the pure preparation. For the proposed pair,
let $\mathcal E_t=e^{t\mathcal L}$ and evolve both pure preparations
under the same reported generator for $t_{\rm pre}=0.04/\Omega_1$.
Their pre-evolved states are
$\bar\rho_X=\mathcal E_{t_{\rm pre}}(|s_X\rangle\langle s_X|)$.
We then set $\rho_X(\tau)=\mathcal E_{\tau/\Omega_1}(\bar\rho_X)$, so
$\tau=0$ begins after that common pre-evolution. After Hermitization
$\rho\mapsto(\rho+\rho^\dagger)/2$ and trace normalization,
double-precision diagonalization gives
\begin{align}
 \operatorname{spec}(\bar\rho_A)&=(1.888285 \times 10^{-7},
  2.978542 \times 10^{-3},0.997021270),\label{SM(resetspectraA)}\\
 \operatorname{spec}(\bar\rho_B)&=(1.677965 \times 10^{-7},
  3.222612 \times 10^{-3},0.996777220).
 \label{SM(resetspectra)}
\end{align}
Both minima exceed the recorded positivity tolerance $10^{-12}$.

The Gorini--Kossakowski--Sudarshan--Lindblad equation \cite{Gorini1976,Lindblad1976} is
\begin{equation}
 \dot\rho=-i[H,\rho]+\sum_{\alpha=1}^2\left(J_\alpha\rho J_\alpha^\dagger-\frac12\{J_\alpha^\dagger J_\alpha,\rho\}\right),
 \label{SM(lindblad)}
\end{equation}
with Eqs.~\eqref{eq:model-H} and \eqref{eq:model}. Vectorization uses column-major order,
\begin{equation}
 \operatorname{vec}(A\rho B)=(B^{\mathsf T}\otimes A)\operatorname{vec}(\rho).
 \label{SM(vectorization)}
\end{equation}
The reported ideal generator gives the following stationary-state eigenvalues:
\begin{equation}
 \operatorname{spec}(\steady)=(0.01837182,\ 0.35519125,\ 0.62643693),
 \label{SM(stationaryspec)}
\end{equation}
The slow decay rate is $1.396755 \times 10^{-2}\,\Omega_1$. The Hermitian left slow mode has eigenvalues
\begin{equation}
 (-0.44300647,\ -0.42561555,\ 0.78904795),
 \label{SM(leftspec)}
\end{equation}
We select the preparation vectors $|\phi_-\rangle$ and $|\phi_+\rangle$ from this mode. In the physical basis $(|0\rangle,|1\rangle,|2\rangle)$, choose global phases so that each first component is real and positive. The vectors used in the calculation are
\begin{align}
 |\phi_-\rangle
 &=\bigl(0.912913205462616,\,
 0.384771520266479\,i,\,
 0.136162977654698\,i\bigr)^{\mathsf T},
 \label{SM(preparation-minus)}\\
 |\phi_+\rangle
 &=\bigl(0.121727563569882,\,
 0.061755770550879\,i,\,
 -0.990640512532678\,i\bigr)^{\mathsf T}.
 \label{SM(preparation-plus)}
\end{align}

For the pair reported in Ref.~\cite{Xia2026}, the kernel crossing times are
\begin{equation}
\begin{array}{c|cc}
\lambda & \tau_- & \tau_+\\ \hline
0.17 & 0.732776 & 0.997354\\
0.25 & 0.447458 & 1.386048\\
0.50 & 0.324503 & 1.942549\\
0.75 & 0.457640 & 2.357716\\
1.00 & 1.339538 & 2.934586
\end{array}
\label{SM(crossingtable)}
\end{equation}
For $\lambda=0.10$, the floating-point calculation found no zero on the displayed window $0.02\le\tau\le3.4$. A longer scan to $\tau=16$ found none either. These finite-window scans do not exclude roots by interval proof. Solving $\Delta_\lambda(\tau)=\partial_\tau\Delta_\lambda(\tau)=0$ gives the numerical branch-merger candidate
\begin{equation}
 (\lambda_c,\tau_c)=(0.1631980171,\ 0.8609542823).
 \label{SM(foldvalue)}
\end{equation}
For the proposed pair, the floating-point search refines every extremum bracketed by the retained $\lambda$ grid and checks both endpoints. The absolute parameter tolerance is $10^{-13}$. Time roots use bracketed scalar solves with absolute tolerance $10^{-12}$. The estimated extrema are
\begin{equation}
\begin{array}{c|cc}
\tau & \widetilde m(\tau) & \widetilde M(\tau)\\ \hline
0 & 5.068964 \times 10^{-2} & 1.7844 \times 10^{3}\\
1.32 & -2.165525 & -5.236941 \times 10^{-2}\\
14.54 & 4.70494 \times 10^{-3} & 1.385378 \times 10^{-2}
\end{array}
\label{SM(margintable)}
\end{equation}
The table rounds the estimates for presentation. Figure~\ref{fig:universal} uses the higher-precision recorded anchor values
$\widetilde m(0)=5.068963644 \times 10^{-2}$, $\widetilde M(1.32)=-5.236940945 \times 10^{-2}$, and
$\widetilde m(14.54)=4.704936829 \times 10^{-3}$.
On $0\le\tau\le2$, the lower-envelope optimizer first leaves
$\lambda=0$ for an interior minimizer at $\tau=0.625667901$, where
$\lambda_{\min}\simeq0.07874$. It returns to $\lambda=0$ at
$\tau=0.709576338$, where the interior branch has
$\lambda_{\min}\simeq0.07038$. The $\lambda=0$ cut has four
floating-point roots at $\tau=0.013320710067$, $0.628770670471$,
$0.704254755999$, and $1.991469435874$. The additional pair survives
only in a narrow layer near the endpoint, and the earliest endpoint
root is numerically ill-conditioned.

Two numerical routes agree: independent eigendecomposition and
matrix-exponential propagation. The Nussbaum--Szko\l{}a expression
also agrees with the direct endpoint evaluation
$\Tr(\rho^{-1}\steady^2)-1$. These are floating-point checks, not
interval bounds. The large upper endpoint value at $\tau=0$ comes
from proximity to a rank-deficient preparation and does not affect
the smallest estimated margin.

The rational-polynomial and grid constructions above provide routes
to interval validation of the ideal model. An experimental validation
would also need a simultaneous confidence region and the conditioning
bound in Eq.~\eqref{SM(profileconditioning)}. The retained numerical
data include every direct cut and field array, additional envelope and
five-kernel arrays, and independently solved zero coordinates. A clean
run reproduced the arrays and scalar values to the stated tolerances.
Outward-rounded bounds and experimental uncertainty propagation remain
separate requirements.

\end{document}